\documentclass[twocolumn,prb,showpacs,preprintnumbers,amsmath,amssymb,floatfix]{revtex4-1}

\usepackage{graphicx}
\usepackage{dcolumn}
\usepackage{bm}

\begin{document}


\title{Dynamics of Dzyaloshinskii domain walls in ultrathin magnetic films
}

\author{Andr{\'{e}} Thiaville}
\author{Stanislas Rohart} 
\affiliation{Laboratoire de Physique des Solides, Univ. Paris-Sud, CNRS UMR 8502,  
91405 Orsay Cedex, France}
\author{{\'{E}}milie Ju{\'{e}}} 
\affiliation{SPINTEC, UMR 8191, CEA/CNRS/UJF/Grenoble-INPG, INAC, 38054 
Grenoble Cedex, France}
\author{Vincent Cros} 
\author{Albert Fert}
\affiliation{Unit{\'{e}} Mixte de Physique CNRS-Thales and Univ. Paris-Sud, 
91767 Palaiseau, France}


\begin{abstract}
We explore a new type of domain wall structure in ultrathin films with
perpendicular anisotropy, that is influenced by the Dzyaloshinskii-Moriya
interaction due to the adjacent layers.
This study is performed by numerical and analytical micromagnetics.
We show that these walls can behave like N{\'{e}}el walls with very high stability,
moving in stationary conditions at large velocities under large fields.
We discuss the relevance of such walls, that we propose to call
Dzyaloshinskii domain walls, for current-driven domain wall motion under
the spin Hall effect.
\end{abstract}


\maketitle

\section{Introduction}
\label{sec:intro}

Ultrathin magnetic films where the magnetic easy axis reorients to the film normal
are one of the first systems in which the specificities of Nanomagnetism became
visible.
A surface contribution to the magnetic anisotropy of a sample was predicted 
early by L.~N{\'{e}}el \cite{Neel54}, and much later observed
to dominate over the volume terms (anisotropy and magnetostatic) in films of 
thickness below 2~nm, typically, when such samples could be fabricated.

As the magnetic degrees of freedom along the film normal correspond to extremely
high energy modes, such samples were assimilated to the practical realization of
the model system in which the simplest magnetic domain wall (DW) exists, namely
the Bloch wall \cite{Bloch32,Landau35}.
In this wall, the magnetization rotates as a spiral when one travels perpendicularly
(direction $x$) across the DW, avoiding magnetostatic charges as 
$\mathrm{div} \vec{m} = d m_x / dx = 0$, where $\vec{m}$ is the unit vector defining the
magnetization direction.
The DW profile is thus governed by the exchange and anisotropy energies, the latter
being in the ultrathin limit an effective anisotropy incorporating the 
perpendicular demagnetizing energy.
As the typical DW widths are of a few nanometers, observations that the DW in such samples
are of this type are extremely rare: spin-polarized scanning tunneling microscopy
would require to control the orientation of the magnetic moment of the last atom, for example.
The most convincing proof was recently obtained in nanostrips of (Co/Ni) multilayer
by anisotropic magnetoresistance measurements, where for small nanostrip widths the DW 
structure was observed \cite{Koyama11} to transform into 
a N{\'{e}}el wall (NW), as expected from micromagnetics \cite{Jung08}.
In the NW, the magnetization rotates in a cycloidal mode, i.e. with the DW normal
$x$ contained in the magnetization rotation plane.
For perpendicular magnetization films that are not patterned into nanostrips, the BW 
does not have a magnetostatic energy but the NW does, so that indeed the BW is the 
lowest energy DW structure, justifying the common belief.

However, it was soon recognized that the energy density difference between NW and BW
decreases as the film becomes thinner \cite{Tarasenko98}, because the `demagnetizing factor'
of the NW reduces as the ratio $t \ln 2/(\pi \Delta)$ of film thickness $t$ to DW width
$\pi \Delta$.
A practical consequence of this is the low Walker field and maximum velocity of the
BW in these films \cite{Tarasenko98}.
Indeed, the stability of a DW structure against precession around a field applied along
the magnetization of the domains is what limits the stationary regime of DW motion,
where the largest DW mobility (velocity to field ratio) and velocity are obtained.
Another direct proof of this low stability of the BW in ultrathin films was recently obtained
through the observation, by ballistic electron emission microscopy \cite{Bellec10},
that in coupled films superimposed walls adopt an antiparallel NW structure.
The stray field from the domains in one layer, despite its smallness because of the
ultrathin film thickness, indeed overcomes the NW demagnetizing field in the other layer.

All these considerations rest on energy terms that are usually considered in micromagnetics.
Recently, another term has been experimentally uncovered in the magnetism of mono- and 
bilayers, namely the Dzyaloshinskii-Moriya interaction (DMI)
\cite{Bode07,Ferriani08,Heide08,Meckler09,Heinze11}.
This antisymmetric exchange interaction was introduced for low-symmetry crystals
\cite{Dzialoshinskii57,Moriya60}.
It was shown to favor non-uniform magnetic structures \cite{Dzyaloshinskii65},
prominently the nowadays so-called skyrmions.
In the context of ultrathin films, A.~Fert first mentioned that, the symmetry being reduced at 
the interface, such a term was also allowed \cite{Fert90}.
This was confirmed by an extensive analytical study \cite{Crepieux98} that considered
the 2-site \cite{Moriya60} and 3-site \cite{Fert80} mechanisms.
Quantitative calculations of this interaction by ab initio techniques 
\cite{Heide06} confirmed its importance for ultrathin films.
Starting with the case of 2 monolayers Fe epitaxially grown on the W(110) surface of a single
crystal, studied in depth by spin-polarized
scanning tunneling microscopy \cite{Bode03}, it was shown that the DMI changes the nature
of the magnetic DW, forcing the existence of N{\'{e}}el walls \cite{Heide08}.
The DW structure has also been investigated for general values of the parameters
and orientation of the Dzyaloshinskii vector \cite{Heide11}, but in a simplified model that
does not solve the magnetostatic problem inherent to a N{\'{e}}el wall.
Moreover, the practical consequences of the introduction of the DMI on the DW dynamics 
have not been investigated so far, with the exception of one work \cite{Tretiakov10}
for a soft magnetic material with a DMI of bulk symmetry.

In this paper, we consider an ultrathin film with perpendicular easy axis, 
grown on a substrate with a capping from a different material
so that the structural inversion symmetry is broken along the film normal ($z$ axis),
the typical sample being Pt/Co(0.6~nm)/AlOx \cite{Miron11}.
The film is patterned into a strip, elongated along the $x$ axis, of a width below the
micrometer but not as small as to induce by magnetostatics the transition to a
N{\'{e}}el wall \cite{Jung08,Koyama11}, providing a well-defined geometry to study the
DW propagation.
As the film is ultrathin, only the $x$-$y$ degrees of freedom exist.
The conventional micromagnetic energy density of the system is
\begin{eqnarray}
\label{eq:E-mumag}
{\cal{E}} &=& A 
\left[(\frac{\partial \vec{m}}{\partial x})^2 +(\frac{\partial \vec{m}}{\partial y})^2 \right]
+ K_\mathrm{u} \left[ m_x^2 + m_y^2 \right] \nonumber \\
&-& \mu_0 M_\mathrm{s} \vec{m} \cdot \vec{H}_a
-\frac{1}{2} \mu_0 M_\mathrm{s} \vec{m} \cdot \vec{H}_d.
\end{eqnarray}
In this expression, $A$ is the (isotropic) exchange constant, $K_\mathrm{u}$ the uniaxial 
anisotropy constant with $z$ the easy axis (sum of volume and surface anisotropies),
$M_\mathrm{s}$ the spontaneous magnetization, $\vec{H}_a$ is the applied field and 
$\vec{H}_d$ the demagnetizing field computed from the magnetization distribution 
through the magnetostatics equations \cite{Hubert98}.
In this geometry, we consider an interface DMI that reads in continuous form
\begin{eqnarray}
\label{eq:E-DM}
{\cal{E}}_\mathrm{DM} &=& D \left[ m_z \frac{\partial m_x}{\partial x} -
 m_x \frac{\partial m_z}{\partial x} \right] +
 \mathrm{id.} \left( x \rightarrow y \right) \nonumber \\ 
&=& D \left[ m_z \; \mathrm{div} \vec{m} - 
\left( \vec{m} \cdot \vec{\nabla} \right) m_z \right].
\end{eqnarray}
This form corresponds to a sample isotropic in the plane, where the Dzyaloshinskii vector
for any in-plane direction $\vec{u}$ is $D \hat{z} \times \vec{u}$ with a uniform constant
$D$, originating from the symmetry breaking at the $z$ surface 
\cite{Fert90,Bogdanov01}.
This interface DMI differs from the bulk-like form where 
$\vec{D}\left[ \vec{u} \right] // \vec{u}$.
The expression (\ref{eq:E-DM}) shows that indeed the DMI favors NWs of a given chirality,
fixed by the sign of $D$.

The previous studies of the DW structure under interface DMI \cite{Heide08,Heide11}
have only considered the statics ($\vec{H}_a = \vec{0}$) and neglected the complex effects
of the demagnetizing field: either $\vec{H}_d = \vec{0}$ or a global demagnetizing effect
was incorporated into an effective anisotropy.
Here, we lift these two restrictions.
Micromagnetic calculations are performed either in 2D (the full problem), or in a model reduced
to 1D (along the $x$ axis) but with a calculation of the demagnetizing field.
We also show how part of the physics of this DW can be captured using the well-known 
collective coordinates
approach (the so-called $q-\Phi$ model), with suitable modifications.
Although most of the paper deals with field-driven motion of DW under DMI in ultrathin films, 
we finally discuss the relevance of the DMI to current-driven DW motion.

The micromagnetic parameters chosen for the simulations are those of the typical sample
considered, namely $M_\mathrm{s} = 1100$~kA/m, $A= 16$~pJ/m, $K_\mathrm{u} = 1.27$~MJ/m$^3$.
This leads to a DW width parameter $\Delta = \sqrt{A/K_0}= 5.65$~nm where
$K_0=K_\mathrm{u}-\mu_0 M_\mathrm{s}^2/2$ is the effective anisotropy including the 
perpendicular demagnetizing field effect in the local approximation.
For the magnetization dynamics, we use the gyromagnetic ratio of the free electron
$\gamma_0 = 2.21\times 10^5$~m/(A.s), and the damping factor $\alpha= 0.5$
that is the typical experimental value \cite{Metaxas07}.

\section{One-dimensional model}
\label{sec:1D}

In a 1D model, the degrees of freedom transverse to the strip are quenched.
As a result, the $y$ derivatives in Eqs.~(\ref{eq:E-mumag}), (\ref{eq:E-DM})
disappear.
Given the typical strip-width ($w=500$~nm) compared to the sample thickness 
($t=0.6$~nm),
we decide to neglect the $y$ component of the demagnetizing field.
In a first approximation, justified by the fact the nanostrip and DW widths
are much larger than the sample thickness, we express the normal component of
the demagnetizing field by the local approximation $H_{dz} = -M_\mathrm{s} m_z$.
The longitudinal component of the demagnetizing field is calculated by direct
summation with interaction coefficients $C_x$ analytically calculated
%
(note also that the $z$ component of the demagnetizing field can be calculated 
similarly, with coefficients $C_z$ applied to $m_z$).
As these interaction coefficients depend on the strip width $w$, this value has to
be specified, even if its influence is very small. 
For the numerical results shown here, it was $w = 500$~nm.
The numerical mesh size was 1~nm, and the length (in the $x$ direction) of
the calculation box was 300~nm.
The box was shifted during the dynamic calculations in order to remain centered
on the DW.

\begin{figure}
\includegraphics[width=0.5\textwidth]{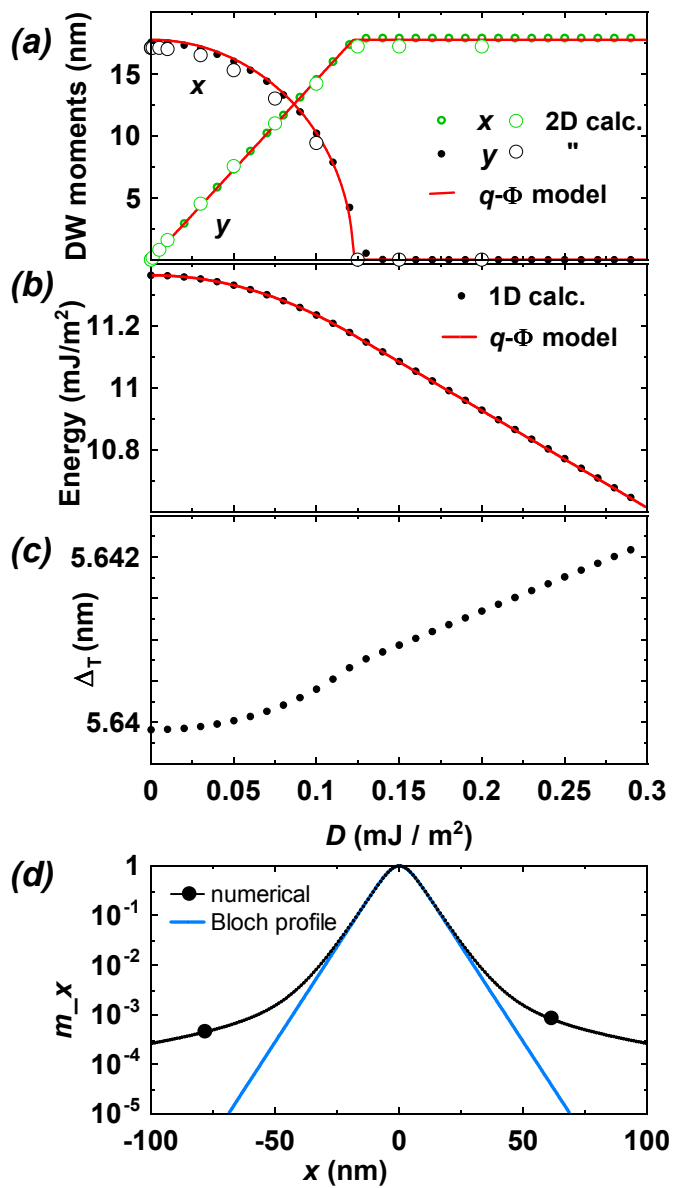}
\caption{
\label{fig:1D-stat}
Equilibrium DW structure according to the 1D model as a function of
the magnitude of the DMI parameter $D$.
(a) integrated DW magnetic moments for the $x$ (N{\'{e}}el) and $y$ (Bloch)
components (big dots show the results of 2D calculation);
(b) DW energy;
(c) DW width according to the Thiele definition.
The magnetization profile is plotted in (d) for the case $D=0.2$~mJ/m$^2$; it 
shows very little variation for larger values of $D$.
Curves in (a,b) were obtained with the $q-\Phi$ model.
}
\end{figure}

The main characteristics of the converged DW structures, as a function of the 
DMI parameter $D$, are shown in Fig.~\ref{fig:1D-stat}.
The two DW moments (the integrated in-plane magnetization components, in nm) 
are plotted in Fig.~\ref{fig:1D-stat}a.
They show that the transition from the BW at $D=0$ to the NW at 
$D > D_c \approx 0.13$~mJ/m$^2$ is progressive.
As we shall see below, this is an effect of the competition between the
DMI and the magnetostatic energies.
The variation of the DW energy, plotted in Fig.~\ref{fig:1D-stat}b, conforms
with the analytic expectation $E = E_0 - \pi D$ \cite{Heide08} at large $D$.
Extrapolating the DW energy to zero, we see that this sample remains in the up-down
state with domain walls up to $D_\mathrm{max} \approx 3.7$~mJ/m$^2$, above which
the nonuniform state develops.

Fig.~\ref{fig:1D-stat}c shows the variation of the DW width parameter
$\Delta_\mathrm{T}$, using the Thiele definition of this parameter that directly 
applies to the DW dynamics \cite{Nakatani05}.
This parameter varies weakly, but the variation is a hint that an analysis
with a Bloch wall profile is too simple.
In order to see this, the profile of the $x$ (N{\'{e}}el) magnetization component 
is plotted in Fig.~\ref{fig:1D-stat}d, and compared to the Bloch wall profile 
with this parameter $\Delta$.
The log scale stresses that, although the Bloch wall profile is very accurate
around the center of the wall, it does not describe the so-called tails that
extend far from the DW.
These tails, well-known in the in-plane case \cite{Hubert98}, are a direct
consequence of the magnetostatic field inherent to the NW structure.
Here, they are responsible for the increase of the wall width parameter 
$\Delta_\mathrm{T}$ as the DW transforms to the NW, opposite to the naive 
expectations based only on the DW core energies.
%

%
\begin{figure}
\includegraphics[width=0.5\textwidth]{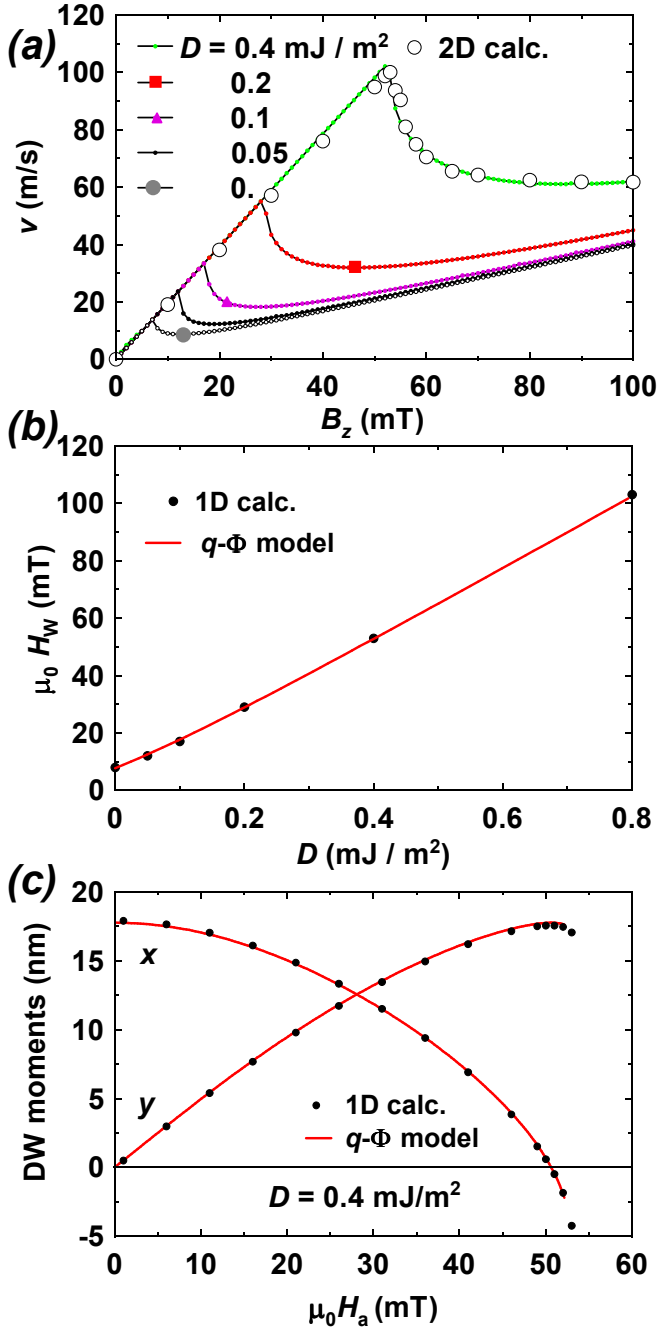}
\caption{
\label{fig:1D-dyn-z}
Field-induced domain wall motion for different value of the DMI
parameter $D$, as calculated by the continuous 1D model.
(a) Velocity-field curves.
The results of the 2D calculation for a width $w= 500$~nm and
$D= 0.4$~mJ/m$^2$ are included for comparison (big dots).
(b) Variation of the Walker field $H_\mathrm{W}$ with $D$.
(c) Evolution of the two DW magnetic moments with field, in the
stationary regime for $D= 0.4$~mJ/m$^2$. 
The curves in (b,c) show the results of the $q-\Phi$ model.
}
\end{figure}

We now study 
the dynamics of these DW structures under a field $H_\mathrm{a}$
applied along the easy ($z$) axis.
As usual, a stationary motion with a high mobility is found at low field, 
followed by a lower velocity regime with a region of negative mobility just
above the critical field called Walker field.
Fig.~\ref{fig:1D-dyn-z}a shows the $v(H_\mathrm{a})$ curves computed for various values
of the DMI micromagnetic parameter $D$ (in the precessional regime, as the
changes are periodic, the average velocity is shown).
The main effect of DMI is to extend the stationary regime, with 
no change of mobility.
The curves are very similar to those of the standard $q-\Phi$
model for Bloch walls \cite{Schryer74}.
In fact, comparing to the curves of that model with identical Walker field, 
Walker velocity and damping, some differences can be seen only in the intermediate 
field region with marked curvature.
The value of the Walker field, plotted in Fig.~\ref{fig:1D-dyn-z}b, shows a quasi
linear increase with $D$.

The quantitative effect is very large: a value $D= 0.6$~mJ/m$^2$ increases
$H_\mathrm{W}$ by ten times compared to $D=0$.
The Walker velocity follows exactly the same trend, simply because the mobility
in the stationary regime is constant (1.96~m/(s.mT)).
Such a strong effect is physically intuitive: the DMI pins the DW magnetization
to the N{\'{e}}el orientation, suppressing the precession of DW magnetization and
thus extending the stationary regime to higher fields.

The robustness of the above results was tested by comparison to exact, but more
demanding, 2D calculations (with suitably modified homemade \cite{Miltat07} and 
OOMMF \cite{OOMMF} codes), 
performed for different nanostrip widths $w \le 500$~nm
and $D < 0.5$~mJ/m$^2$.
As soon as $w$ was larger than the BW-NW magnetostatic transition \cite{Jung08,Koyama11},
the evolution of the results with $w$ was minor.
Neither a continuous rotation of the DW magnetization along the length of 
the wall, nor a tilt of the wall, were observed, so that the 1D model is justified, 
as it describes a stable structure.
All quantities (DW width, Walker field, etc.) were found to be slightly modified, 
mainly because of the assumption of a local $z$ demagnetizing field with $N_z=1$, so
that the effective anisotropy is larger and the DW width lower.
The small variations, observed above the Walker field, are due to
the transverse degrees of freedom (see Ref.~\cite{Yamada11} for an example).
They are more apparent as $w$ or $D$ increase.
In order to give an idea of the magnitude of the 2D effects, one such calculation
result is added in Fig.~\ref{fig:1D-dyn-z}a.

\section{Collective coordinates $q-\Phi$ model}
\label{sec:qphi}

Despite the complexity of the NW profile (Fig.~\ref{fig:1D-stat}d), and of the
local form of the DMI, a simplified model based on only two collective 
coordinates, namely the DW position $q$ and the DW magnetic moment angle
$\Phi$, can be constructed, similarly to what has been done for the
conventional DW \cite{Schryer74}.
The starting point is to express the DW surface
energy $\sigma$ \cite{Slonczewski72}.
As seen above, for the typical system that we consider the DW 
width is practically constant.
We therefore assume that the wall width parameter is a constant
$\Delta = \sqrt{A / K_0}$.
We also make the assumption that the magnetization rotates in a vertical plane, 
rotated by an angle $\Phi$ around the $z$ axis ($\Phi=0$ for the NW).
The 1D calculations show that this is not the case, neither far from the DW
as $m_x$ decays slowly whereas $m_y$ does not, nor around the center of the wall
when driven by a field.
Nevertheless, this assumption allows a crucial simplification and the final results
match quite well with 1D calculations, as shown below.
In these conditions, the DW energy reads
\begin{equation}
\sigma = 2 \Delta K \cos^2 \Phi - \pi D \cos \Phi 
 + C^\mathrm{st},
\label{eq:sigma}
\end{equation}
where $K$ is the magnetostatic `shape' anisotropy that favors the Bloch wall,
related to the `demagnetizing coefficient' $N_x$ of the wall by
$K = N_x \mu_0 M_\mathrm{s}^2/2$.
The Slonczewski equations derived from Eq.~(\ref{eq:sigma}) are then,
with $H_K = 2 K /(\mu_0 M_\mathrm{s})$ and
$H_D = \pi D / \left( 2 \mu_0 M_\mathrm{s} \Delta \right)$,
\begin{eqnarray}
\label{eq:Slonc1}
\dot{\Phi} + \alpha \dot{q} / \Delta &=& \gamma_0 H_\mathrm{a} \\
\dot{q}/\Delta - \alpha \dot{\Phi} &=& \gamma_0 
\left( -H_K \sin\Phi \cos\Phi + H_D \sin\Phi \right)
\label{eq:Slonc2}
\end{eqnarray}

Minimizing $\sigma$ with respect to $\Phi$ at $H_\mathrm{a}=0$ solves the statics.
The stable solution is
\begin{eqnarray}
\label{eq:equil}
\cos\Phi_0 &=& \pi D / \left( 4 \Delta K \right) \;\mathrm{for}\; \pi |D| < 4 \Delta K 
\\ \nonumber
&=& \mathrm{sign}(D) \;\mathrm{for}\; \pi |D| > 4 \Delta K.
\end{eqnarray}
Thus a critical value $D_\mathrm{c} = 4 \Delta K / \pi$ appears.
When $|D| > D_\mathrm{c}$ we have a NW, and below this limit the DW moment 
reorients smoothly to the Bloch orientation.
Figs.~\ref{fig:1D-stat}a,b show the good overall agreement of this model
with the numerical results of the 1D model (for all calculations with
the $q-\Phi$ model, we used $\Delta= 5.64$~nm and $N_x = 0.0224$, derived
from the 1D calculations).

The model is also analytically solvable in the stationary regime of the
dynamics.
Taking $\Phi$ as a parameter, one has indeed
\begin{equation}
\label{eq:Hdyn}
H_\mathrm{a} = \alpha \sin\Phi \left( H_D - H_K \cos\Phi \right).
\end{equation}
Stability of $\Phi$ requires also that 
$H_K \cos 2 \Phi - H_D \cos\Phi <0$.
The DW velocity $v$ is then given by the stationary relation
$v = \left( \gamma_0 \Delta / \alpha \right) H$, where.
Fig.~\ref{fig:1D-dyn-z}c shows how well this reproduces the numerical
results of the 1D model for the case $D= 0.4$~mJ/m$^2$.
The highest field applicable, the Walker field, is found to
be given by
$\cos\Phi_\mathrm{W} = \left( \delta - \sqrt{\delta^2 + 8} \right) / 4$
where $\delta \equiv D / D_\mathrm{c} = H_D / H_K$
(the expressions for the Walker field $H_\mathrm{W}$ and the
Walker velocity $v_\mathrm{W}$ are cumbersome but straightforward
to write; at large $D$ one has simply 
$H_\mathrm{W} \approx \alpha H_D$).
The comparison with the Walker field data from the numerical
solution of the 1D model is depicted in Fig.~\ref{fig:1D-dyn-z}b,
again with a very good agreement.
In the precessional regime, the dynamic equation for $\Phi$ alone, 
derived from Eqs.~(\ref{eq:Slonc1}),(\ref{eq:Slonc2}), should be 
integrated in order to get the oscillation period as function of field.
Comparison with the curve of the standard model ($D=0$) shows
some difference only in the regime where $H_\mathrm{a}$ is not so much 
above $H_\mathrm{W}$, as may be anticipated.

\section{Comparison to experiments}
\label{sec:expe}

A first hint that the DW in ultrathin Pt/Co (0.6)/AlOx films is not a simple BW 
is provided by the DW dynamics under field only in this sample \cite{Miron11},
compared to those of a series of Pt/Co(t)/Pt with $t=0.5-0.8$~nm
\cite{Metaxas07}: the velocity in the flow regime is indeed at least
5 times larger in the former case (the comparison is not so direct
as the Co/AlOx interface gives more interface anisotropy than the
Co/Pt interface).
The latter study concluded that the flow regime was precessional,
with $\alpha \approx 0.3$, the stationary regime being hidden by
the DW pinning.
As a result, the high mobility observed for Pt/Co/AlOx can only be
obtained in the stationary regime, but in that case the Walker field 
would be smaller than the fields applied (up to 120~mT).
As shown by Fig.~\ref{fig:1D-dyn-z} and by the analytic model,
this paradox can be lifted with $D \approx 1$~mJ/m$^2$ or larger.
This value is reasonable: we estimate for Fe/W(110) 
\cite{Heide08} 4~mJ/m$^2$, for Mn/W(001) \cite{Ferriani08} 3~mJ/m$^2$,
and for Fe/Ir(111) \cite{Heinze11} 1.5~mJ/m$^2$, supposing a dilution
of the DMI values found in mono- or bilayers over a 3 monolayers film.

\begin{figure}
\includegraphics[width=0.5\textwidth]{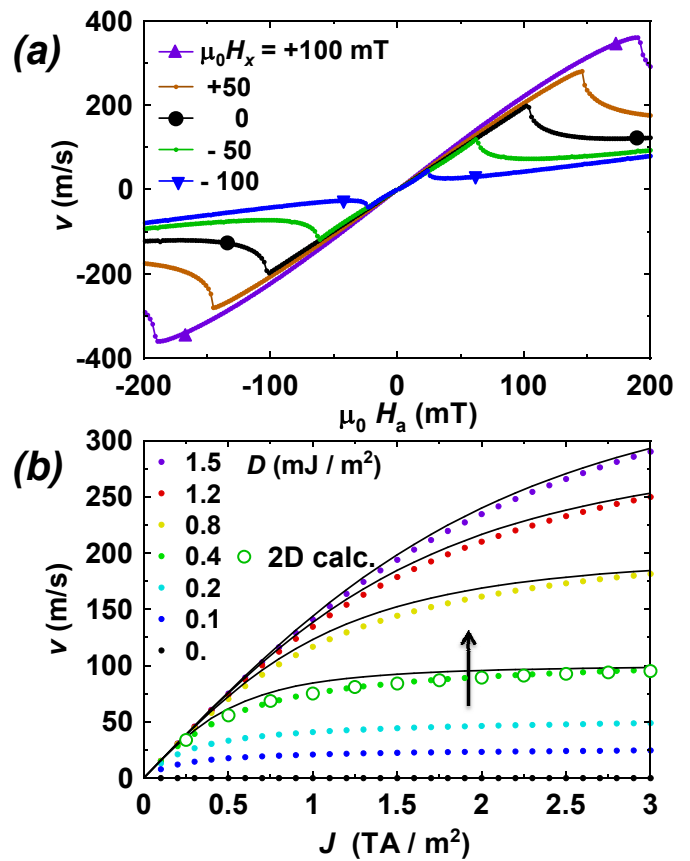}
\caption{
\label{fig:1D-dyn-xJ}
Dynamics of a Dzyaloshinskii DW as calculated by the 1D model:
(a) field-induced motion under an additional in-plane field
along the nanostrip, for $D= 0.8$~mJ/m$^2$;
(b) current-induced motion due to the spin Hall effect in the
underlayer, for a Hall angle $\theta_H=0.1$~rad and several values of $D$,
with results of 2D calculations superposed for one case
(arrow depicts $D$ increase, and curves show the large $D$ solution of the 
extended $q-\Phi$ model). 
}
\end{figure}

One test of the NW structure due to DMI is to measure the
DW dynamics under a $z$ field, with an additional in-plane field
along the $x$ direction.
Indeed, in simple terms, this field favours opposite chiralities
for up-down and down-up walls, competing with the DMI.
The results of the full 1D model for this case are shown in
Fig.~\ref{fig:1D-dyn-xJ}a, for an up-down wall with $D=+0.8$~mJ/m$^2$.
The positive $x$ fields add to the DMI for stabilizing the
DW moment in the $x$ orientation, and increase the Walker field
(and Walker velocity).
In the expression of the DW energy $\sigma$, the $x$ field would
enter like a DMI term, but the magnetization rotation that it
induces in the domains should also be taken into account.
This stresses the main interest of the DMI: it provides a
large $x$ field that exists only within the wall.
The figure thus shows that $D= 0.8$~mJ/m$^2$ is equivalent to 
a stabilizing field $\mu_0 H_x \approx 128$~mT,
which is exactly the value $D/(M_\mathrm{s} \Delta)$ of the average over
the DW of the equivalent $B_x$ field deduced from Eq.~(\ref{eq:E-DM}).
This is very different from the case of a transverse ($y$) field, where
small changes of mobility and Walker field are observed, depending on the
sign of the $z$ field (not shown).

We proceed now to current-induced DW motion in ultrathin films
with perpendicular anisotropy.
Recent work \cite{Khvalkovskiy12,Haazen12}, has revealed that the nature of
the DWs in such samples is very important.
Indeed, to explain the anomalously large DW velocity under current in 
Pt/Co/AlOx, a magnetic effect of the Rashba field at the interfaces has been 
proposed \cite{Miron10b}.
On the other hand, a spin Hall effect (SHE) due to the current flowing in the
Pt or Ta buffer layer has also been considered \cite{Liu12}.
The analysis of the action of these torques \cite{Khvalkovskiy12} shows
that for perpendicular materials the SHE can efficiently move NWs at
zero field and in opposite directions for opposite chiralities.
This has been directly proven in the Pt/Co/Pt case \cite{Haazen12}, by
preparing NW with moderate $x$ fields and observing a current assisted
depinning only for the NW.
In that structure, the sign of the SHE was moreover controlled by the relative
thicknesses of the two Pt layers, and a strong DMI is not expected
as the effects of both interfaces compensate.
We thus predict that, for asymmetric structures where the two interface DMI
do not compensate, the DW may be a NW that efficiently moves under 
current by SHE, without the need of an applied field. 
Its direction of motion with respect to the current direction is given by the 
sign of the DMI parameter and the SHE film position (above or below the magnetic
film).
In addition, as the NWs induced by DMI have a fixed chirality, all
walls will be pushed in the same direction under current by the SHE.
This feature is highly desirable in applications where one wants to obtain
the motion of a train of DW in the same direction (race-track memory
for example).
It cannot be obtained with the opposite successive chiralities induced
by an $x$ field \cite{Haazen12}.

Calculations with the 1D model (Fig.~\ref{fig:1D-dyn-xJ}b) indicate that 
DW velocity under SHE increases with $D$, becoming comparable to 
experiments \cite{Miron11} for $D > 1$~mJ/m$^2$ for a spin Hall angle
of 0.1~rad.
The $q-\Phi$ model can also be modified to include this term.
Using the Thiele force equation approach, we find that the SHE can be input 
by an equivalent field
$H_\mathrm{a} = \left(\pi/2\right) \chi M_\mathrm{s} \cos \Phi$, with 
$\chi = \hbar \theta_\mathrm{H} J / \left( 2 e \mu_0 M_\mathrm{s}^2 t \right)$ 
representing the SHE-induced STT \cite{Stiles06}, 
$J$ being the current density.
In the steady-state regime, and for $\delta \gg 1$ we obtain
$v = v_D / \sqrt{1+\left( J_D / J \right)^2}$, with
$v_D = \gamma_0 \Delta H_D$
and $J_D = 2 \alpha t e D / \left( \hbar \theta_H \Delta \right)$
(note that the $v/J$ slope at $J=0$ is independent of $D$).
The agreement with the results of the full 1D model is good 
(Fig.~\ref{fig:1D-dyn-xJ}b), despite the neglect of the magnetization
rotation in the domains under the SHE.

\section{Conclusion}

We have studied by micromagnetics a new domain wall structure in ultrathin films with 
perpendicular
anisotropy, as a result of uncompensated Dzyaloshinskii-Moriya interactions 
from the adjacent layers, that we propose to call Dzyaloshinskii domain wall (DDW).
Depending on the strength $D$ of this interaction, domain wall magnetizations intermediate 
between the Bloch and N{\'{e}}el orientation are obtained, turning to
N{\'{e}}el at large $D$.
The motion of a DDW under field is strongly affected by $D$, with a nearly linear
increase of Walker field and velocity with $D$, up to very large values
before the domains become unstable.
Specific features of the dynamics of the DDW under in-plane fields
have also been described.
DDWs offer an attractive scenario to contribute to the understanding of the
wealth of experimental results on currrent-induced DW motion in ultrathin films 
with perpendicular anisotropy, and for applications.

\acknowledgments
This work was supported by the Agence Nationale de la Recherche, project
ANR 11 BS10 008 ESPERADO. 
Discussions with J. Vogel, G. Gaudin, S. Pizzini, I.~M. Miron, O. Boulle 
and J. Miltat are gratefully acknowledged.

\end{document}